\documentstyle[sprocl]{article}

\input{psfig}

\def\Journal#1#2#3#4{{#1} {\bf #2}, #3 (#4)}

\def\NPB{{\em Nucl. Phys.} B}
\def\PLB{{\em Phys. Lett.}  B}
\def\PRL{\em Phys. Rev. Lett.}
\def\PRD{{\em Phys. Rev.} D}

\def\be{\begin{equation}}
\def\ee{\end{equation}}
\def\bea{\begin{eqnarray}}
\def\eea{\end{eqnarray}}

\begin{document}

\title{SEARCH FOR QCD INSTANTON-INDUCED PROCESSES 
IN DEEP-INELASTIC SCATTERING AT HERA}

\author{S. MIKOCKI,\\ FOR H1 COLLABORATION}

\address{Institute of Nuclear Physics, Cracow, Poland}


\maketitle\abstracts{ Preliminary H1 results on 
dedicated searches for QCD instanton induced processes 
in DIS are presented. The investigations are based
on the expected  characteristics of their hadronic final state. 
Searches were performed 
in the kinematical region $x > 10^{-3}$ , $0.1 < y < 0.6$ 
and $\theta_{el} >156^o$.}
\section{Introduction}
Instantons~\cite{bel}, non-perturbative fluctuations of 
non-abelian gauge fields, induce  anomalous processes which violate
classical conservation laws like baryon plus lepton number
in the case of the electroweak interactions and chirality  in the case of QCD.
Deep-inelastic scattering (DIS) at HERA offers 
the unique possibility~\cite{sch1} to discover 
processes induced by QCD-instantons. 
 The theory~\cite{sch234} 
and phenomenology~\cite{sch1,sch5}
have been worked out by A. Ringwald and F. Schrempp. 
The  expected event topology can be simulated with   
the QCDINS~\cite{sch6} Monte Carlo program.
The predicted cross section~\cite{sch0}
is  $\sigma_{ins} \approx$ 30 - 100 $\rm pb$
for $0.1 < y < 0.9$ and $x > 10^{-3}$.\footnote{
We use QCDINS2.0 with the default values. We do not apply the 
$Q^{2} > 100~GeV^2$ cut 
to suppress non-planar diagrams as 
recently recommended by the authors.}
%
We can therefore expect a sizeable number of such events.
The background, however, is three orders of magnitude higher.
\section{Data Selection}
The analysis is based on data taken in 1997 with
the H1 detector corresponding to 
an integrated  luminosity  of  $ {\cal L} =15.78~\rm pb^{-1}$.
The analysis is performed in the following phase space: 
  $0.1 < y_{el} < 0.6$,   $x_{el} > 10^{-3}$
and $\theta_{el} > 156^\circ$. The charged particles are reconstructed
  in the acceptance region of the Central Track Chambers
$20^\circ < \theta < 155^\circ$ with the transverse momentum 
$p_T > 0.15$ GeV. A combination of tracks and energy depositions 
in the calorimeter
is used to measure the energy flow.~\cite{h1} 
The total DIS sample contains $\sim 280000$ events.
%
%
%
\section{Observables and Search Strategy}
\begin{figure}[t]
\psfig{figure=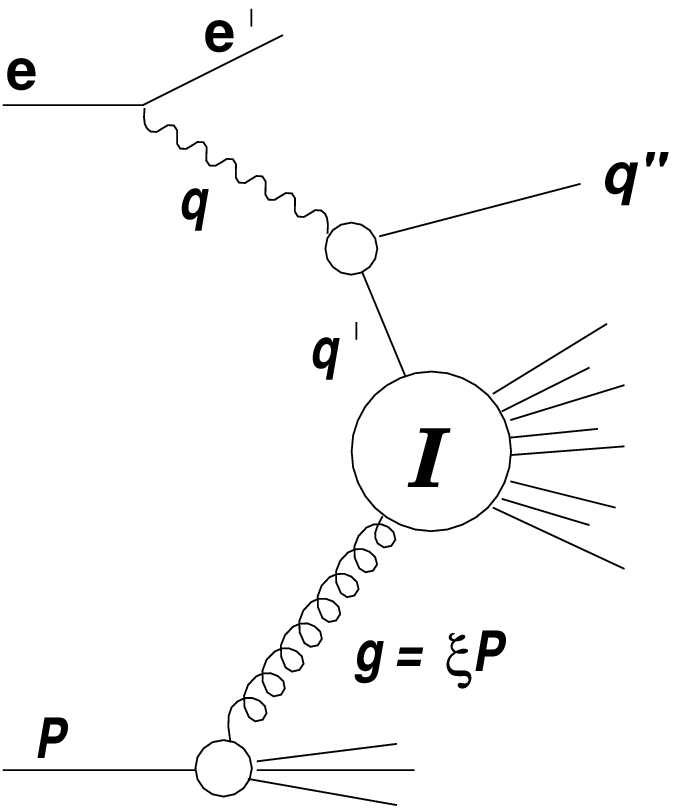,height=2.1in}
\psfig{figure=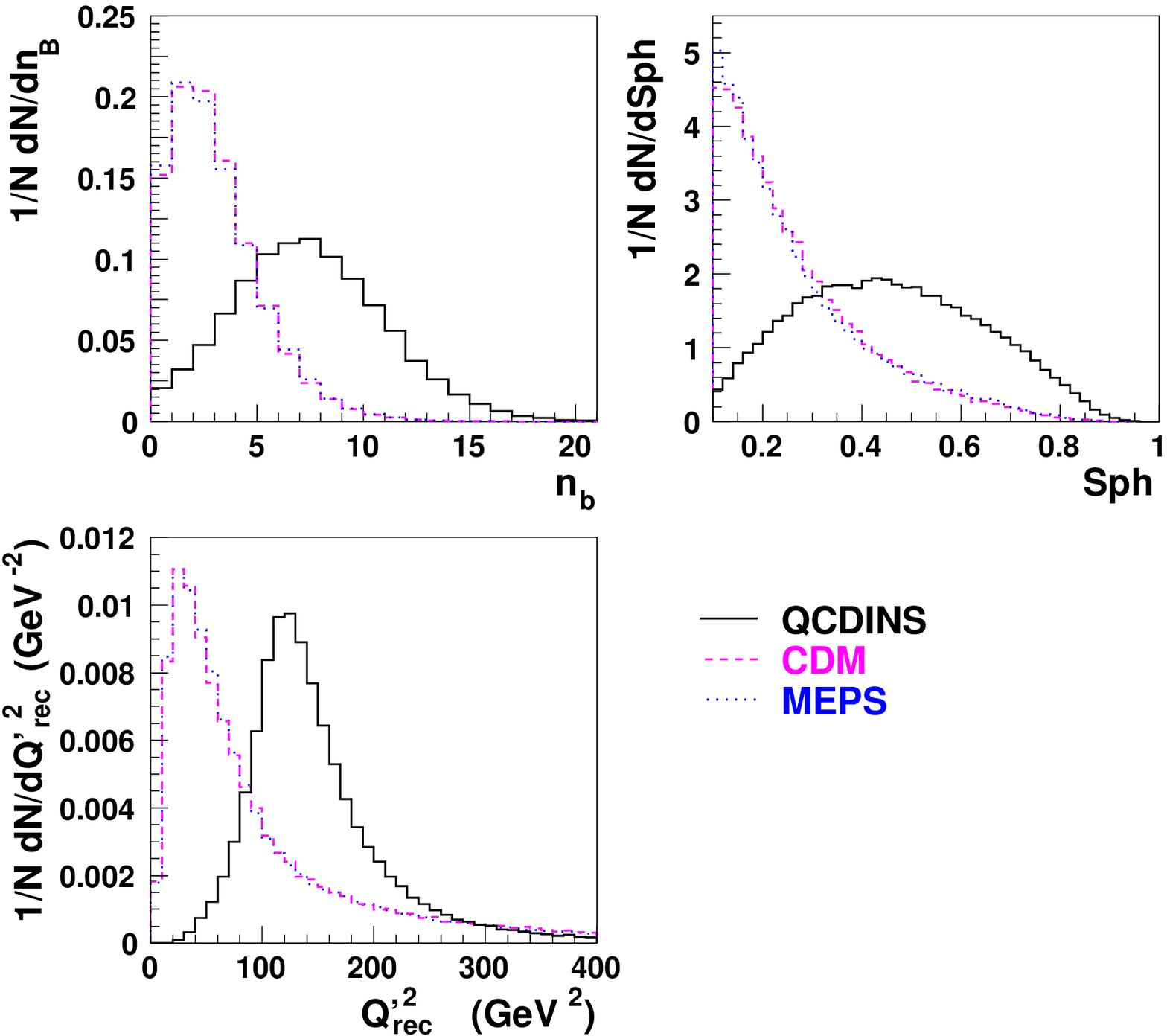,height=2.1in}
\caption{Left: Sketch of an instanton-induced process. Right:
Shape normalised distributions of the  observables used to discriminate
I-processes from normal DIS.
  \label{fig:strategy}}
\end{figure}

 The instanton-induced 
 hadronic final state is expected to have the following  signature
 (see Fig.~\ref{fig:strategy}):  
 a densely populated narrow band in pseudorapidity which is homogeneously
 distributed in azimuth
 (isotropic 'fireball'-like final state),
a large total transverse energy, 
a large particle multiplicity including
all kinds of flavours e.g.  strange particles.
The high background from 
normal DIS events requires the best possible discrimination.
    
The following observables\footnote{All observables are calculated 
in the hadronic CMS ($\vec q + \vec P = \vec 0$) except the sphericity.} have been used
to discriminate I-induced processes from normal DIS events:
(1) $Et_{jet}$, the jet with highest $E_T$ (cone algorithm 
with radius $R=0.5$). We associate   this jet with the 'current'
quark (q") in Fig.~1.   
(2)  The virtuality of the quark entering the I-process 
${Q'}^2 = -(q-q")^2$    
 where the photon (q) is reconstructed by the scattered
electron.
(3)  The number of charged particles $n_B$ in the instanton band.\footnote{
 The particles belonging to the jet $q"$ are removed from the final state
and
the $E_T$-weighted mean pseudorapidity $\bar{\eta}$
is recalculated with the remaining ones.
The instanton band is  defined as ~$\bar{\eta} \pm 1.1$.}          
(4)  The sphericity SPH  calculated in the rest system of the
particles not associated with the current jet. 
(5) $Et_b$  the total transverse energy in the
instanton band calculated  
as the scalar sum of the transverse energies
and (6) $\Delta_b$,\footnote{ $\Delta_b$ is defined as 
$
\Delta_b = \frac{E_{in} - E_{out}}{E_{in}}
$
where $E_{in}$ ($E_{out}$) is the maximal (minimal) value of 
the sum of the projections  on all possible axis $\vec{i}$
of all energies depositions in the band 
(i.e. $E_{in}= max \sum_n |\vec{p_n} \vec{i} |$). 
For isotropic events $\Delta_b$ is expected to be small
 while for jet-like events it  should
be large.} a quantity measuring the $E_T$ weighted $\Phi$ event isotropy. 

To find the optimal cut scenario
the following cut values have been applied:
$
n_B > 5, 6, 7, 8, 9
$
;
$
SPH > 0.4, 0.5, 0.55, 0.6, 0.65
$
and
$95, 100, 105, 110, 115 <  {Q'}^2 < 200\;{\rm GeV}^2$ (see Fig.~\ref{fig:strategy}).
From 125 cut combinations   
three scenarios are chosen according to the following criteria:
(A) The highest instanton efficiency ($\epsilon_{ins}$ $\approx 30 \%$),  
(B) High $\epsilon_{ins}$ at reasonable background reduction and  
(C) Highest background reduction ($\epsilon_{dis}$ $\approx 0.13-0.16 \%$)
      at   $\epsilon_{ins}$ $\approx 10 \%$.
%
%
%
\begin{table}[t]
\caption{Measured numbers of events and expected background
 for three cut scenarios. The errors are dominated by systematic errors.}
\vspace{0.2cm}
\begin{center}
\footnotesize
\begin{tabular}{|c|c|c|c|c|c|}
\hline
\multicolumn{2}{|c|} {\raisebox{0pt}[12pt][6pt]{ (A) DATA: $3000$ }} &
\multicolumn{2}{|c|}{\raisebox{0pt}[12pt][6pt] {(B) DATA: $1332$} } &
\multicolumn{2}{|c|}{\raisebox{0pt}[12pt][6pt] {(C) DATA: $ 549$} }\\
\hline
\raisebox{0pt}[12pt][6pt] {CDM} & {MEPS}  &    
 {CDM} & {MEPS}  & 
 {CDM} & {MEPS}  \\ 
\hline
 \raisebox{0pt}[12pt][6pt] {$2469^{+242}_{-238}$} & {$2572^{+237}_{-222}$} &
  {$1005^{+82}_{-70}$} & {$1084^{+75}_{-46}$} &
  {$ 363^{+22}_{-26}$} & {$435^{+36}_{-22}$} \\
\hline
\end{tabular}
\end{center}
\label{table:result}
\end{table}
\begin{figure}[t]
\psfig{figure=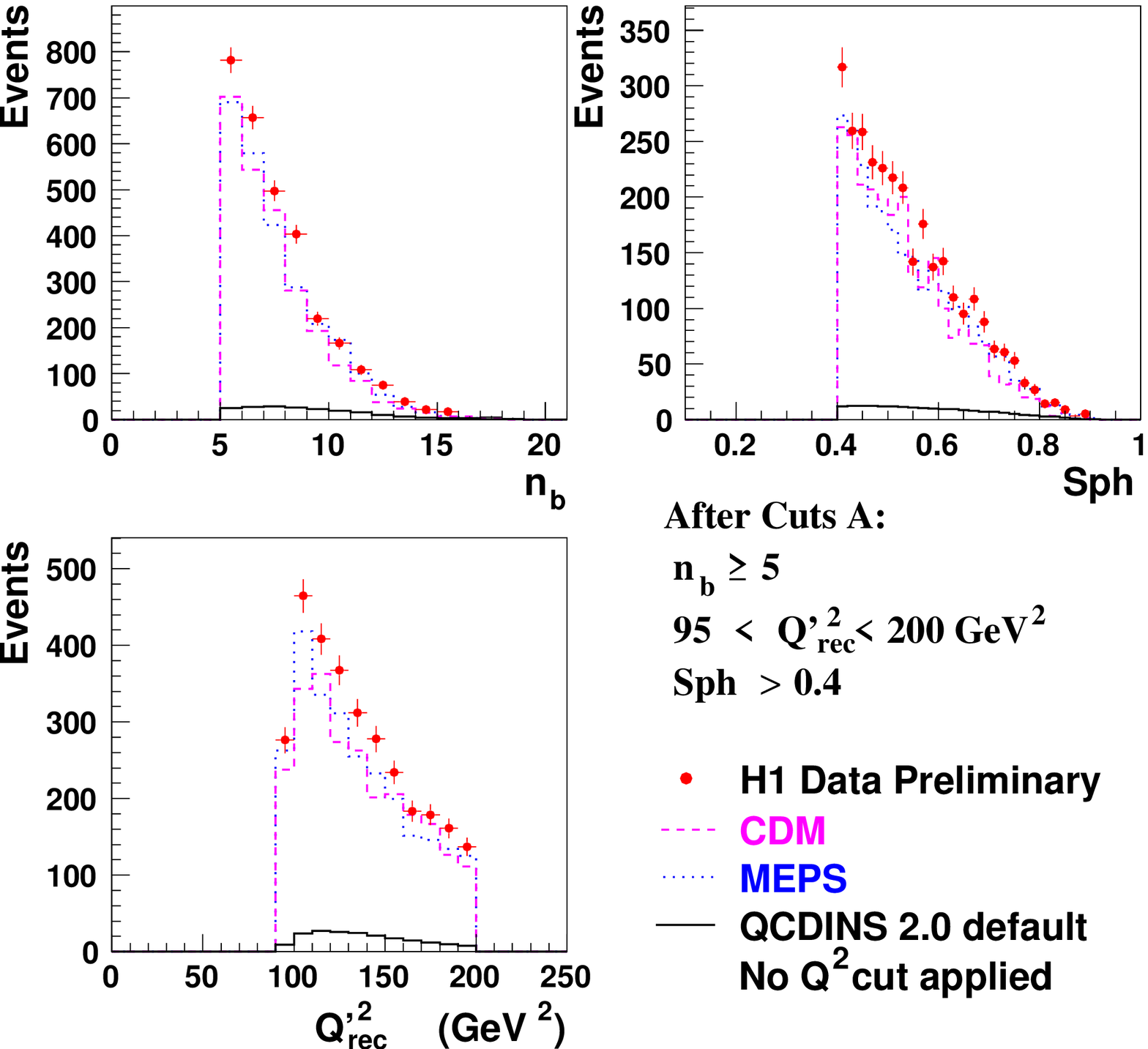,height=2.1in}
\psfig{figure=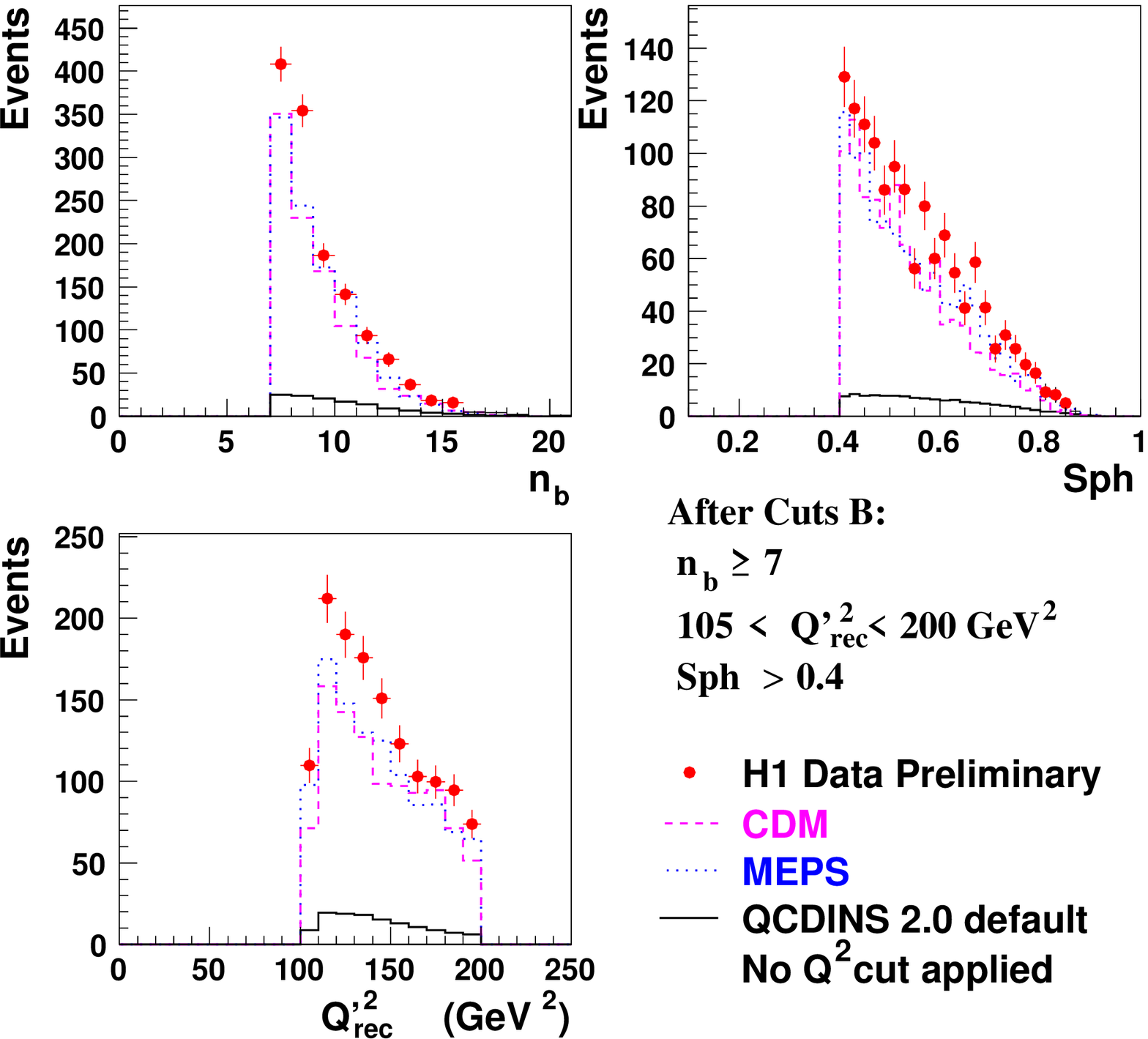,height=2.1in}
\psfig{figure=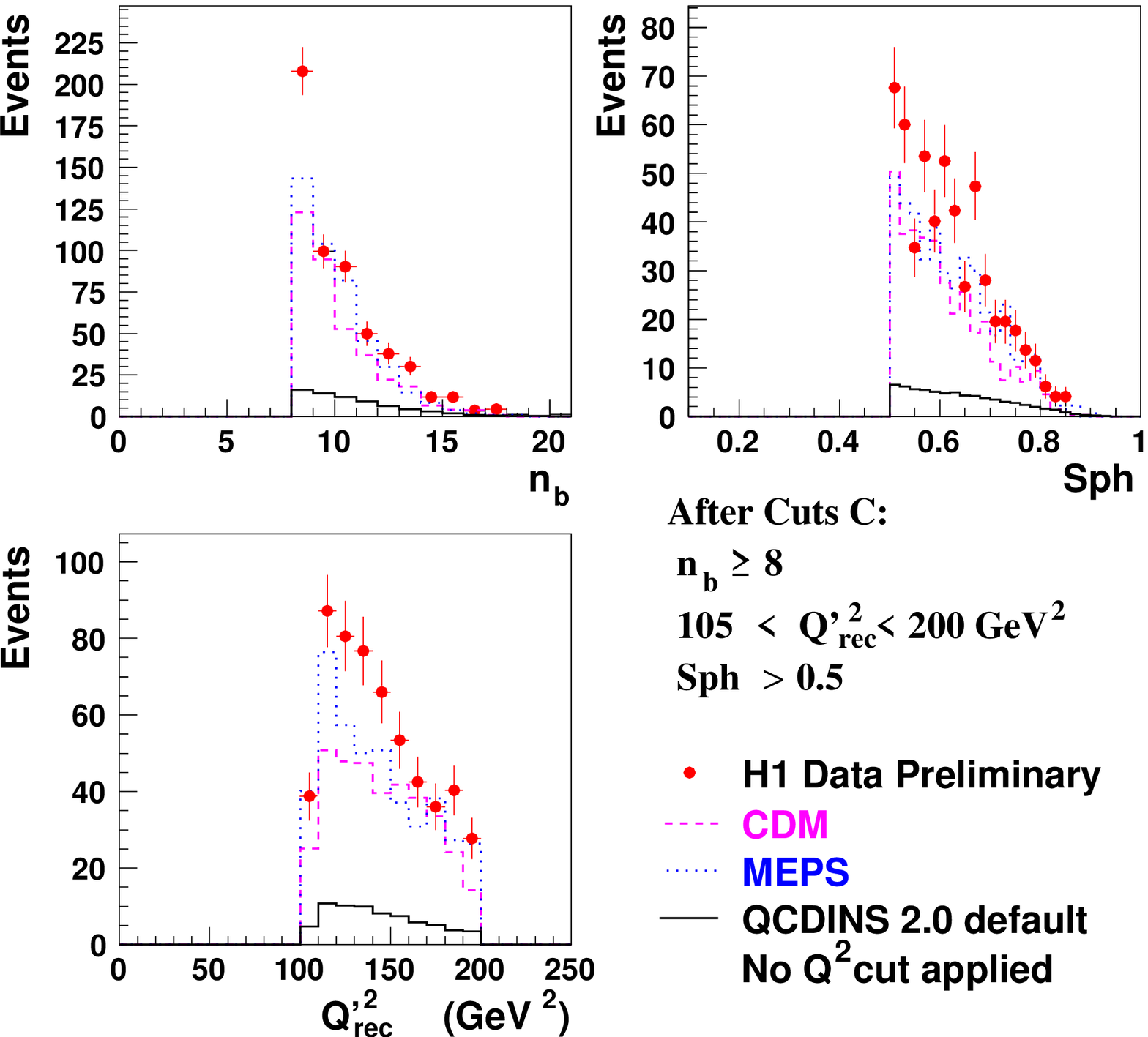,height=2.1in}
\psfig{figure=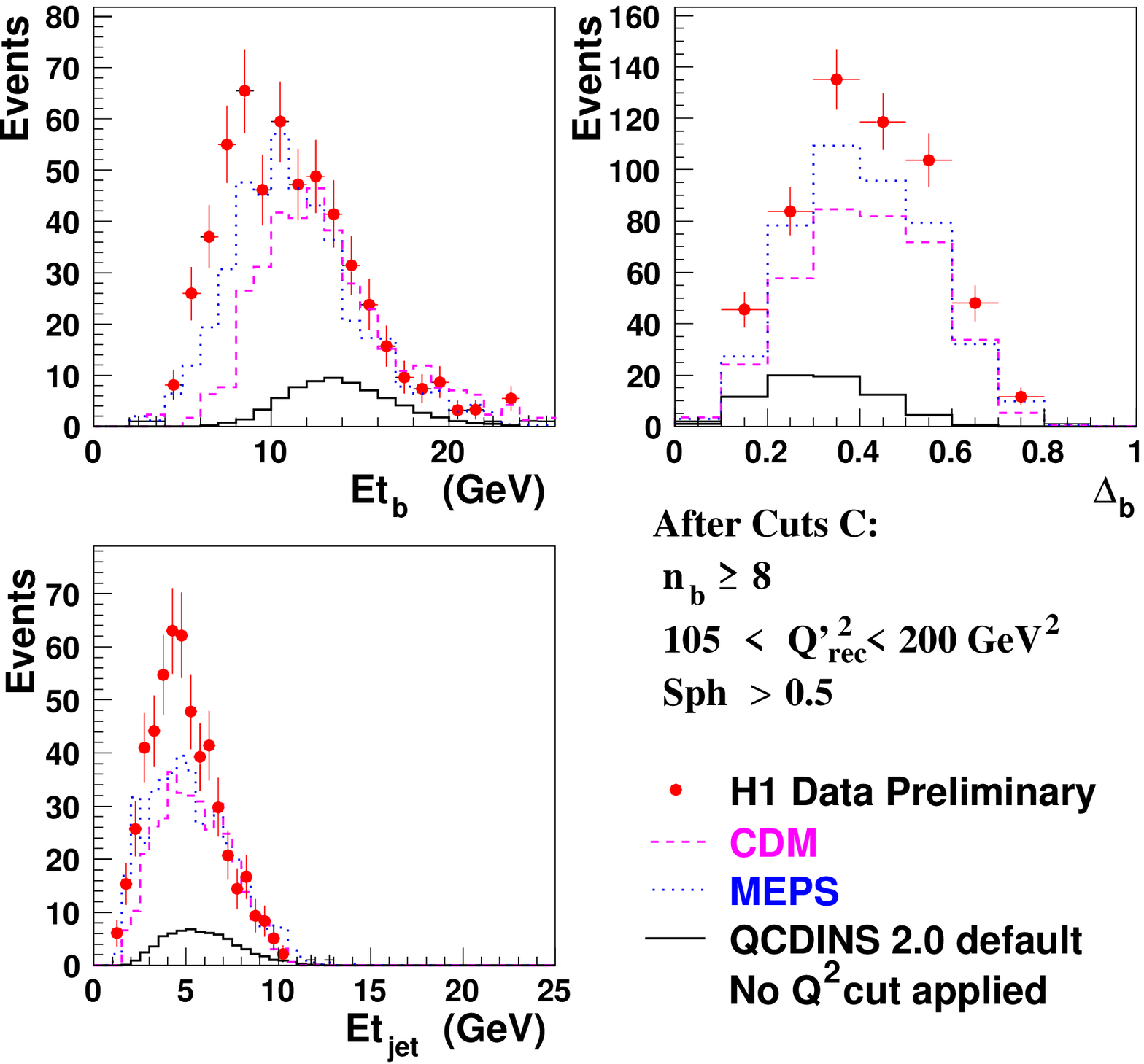,height=2.1in}
\caption{The distributions of the observables used in cuts for three cut scenarios
 and (lower right) the distributions of observables not used in cuts
 for scenario C.
  \label{fig:aftercuty}}
\end{figure}  
\section{Results and Conclusions}
The results are presented in Fig.~\ref{fig:aftercuty} and Table~\ref{table:result}.
In all scenarios more events are  observed than expected by 
the standard QCD models CDM~\cite{ariadne} and MEPS.~\cite{rapgap}
The shape of the excess in $n_B$, SPH and $Q'^{2}$ is qualitatively
compatible with the expected instanton signal. However, the size of this
signal is  at the level of differences between the QCD models. 
The shapes of the other observables (not used in the cuts)
are neither well reproduced by CDM nor by MEPS. The observed excess in $Et_b$
and $\Delta_b$  is  not particularly favoured by the QCDINS
predictions.

QCDINS contributions cannot be excluded from the data given the uncertainties
in their calculation and modelling.~\cite{sch0} In addition, a better
understanding of the DIS hadronic final state formation in the phase
space relevant for instanton searches is required.

\end{document}